# STELLAR SOURCES OF GAMMA-RAY BURSTS

B.I.Luchkov
National Research Nuclear University MEPhI
Moscow, Kashirskay Road 31

Correlation analysis of Swift gamma-ray burst coordinates and nearby star locations (catalog Gliese) reveals 4 coincidences with good angular accuracy. The random probability is $4 \times 10^{-5}$, so evidencing that coincident stars are indeed gamma-ray burst sources. Some additional search of stellar gamma-ray bursts is discussed.

Cosmic gamma-ray bursts (CGB), short duration gamma-ray fluxes with energies 10 keV - 10 MeV registered on satellites and spaceships, were mystery for a long times [1]. The CGB nature was unknown: what are their sources, the distances and processes in which them produced? Some device resolution improvement to about angular minutes came to discovery of source X-ray and optical afterglow. Large ground telescopes and Hubble Space Telescope could find fading transient objects. Measuring in some cases redshift Z established the transient nature [2]. About 40 % of them have $Z \geq 1$ and consequently are on large (cosmological) distances appearing as fireballs – Supernova flares at remote galaxies [3]. However the gamma-burst mystery was not entirely revealed: what are the other 60 % ?

Coming from idea that the CGB sources could present not single class but summery of different objects according their power, location and emitting processes, an assumption was given about near sources [4], among which could be active (flaring) stars of small masses. They are young stars of late spectral classes [5].

This work is devoted to investigation of nearby stars which flares could generate gamma-ray bursts.

**Correlation analysis of gamma-ray burst and star locations**

Gamma-burst positions are taken from Swift catalog [6] which contains now (February 2011) 391 unknown CGBs and 175 cosmological CGBs with $Z > 0.5$. Swift angular accuracy is $\sigma \approx 0.1^\circ$. CGB coordinates are compared with coordinates of nearby stars from Gliese catalog [7, 9]. In order to decrease number of random applications the stars are taken with parallaxes $P > 0.05$ ($r \leq 20$ ps), star value $m < 8$ and only of G, K, M spectral classes which eventually could be gamma-ray burst sources due to their flare activity. Thus number of stars taken for correlation with gamma-bursts was $N_{st} = 470$. The cosmological CGBs which could coincide only randomly serve as a comparison criterion.

As numerical indicator for CGB and star coordinates coincidence a deviation $\Delta r = (\Delta\alpha^2 + \Delta\delta^2)^{1/2}$ was used, where $\Delta\alpha$ and $\Delta\delta$ are angular differences for right ascension and declination. The analysis result is given in table 1.

Table 1. Small deviations $\Delta r$ (degree)

|  | 0-0.12 | 0.12-0.24 | 0.24-0.36 | 0.36-0.48 | 0.48-0.60 | 0.60-0.72 | 0.72-0.84 | 0.84-0.96 |
|---|---|---|---|---|---|---|---|---|
| CGB | 1 | 3 | 4 | 3 | 3 | 4 | 8 | 9 |
| Cosmological CGB (Z) | 0 | 0 | 3 | 1 | 2 | 2 | 3 | 6 |

As one can see 4 coincidences of CGB and stars coordinates were found with $\Delta r < 0.24$

corresponding to Swift angular resolution. There are no similar near coincidences among control group of cosmological CGBs.

Four CGB ($\Delta r = 0.08^o$, $\Delta r = 0.14^o$, $\Delta r = 0.19^o$, $\Delta r = 0.21^o$) are evidently of stellar origin. The important evidence is obtained by strong gamma-ray burst of nearby flaring star EV Lacertae (red rotating dwarf, M3.5 spectral class, at r = 5 pc) [8]. It is right proof of stellar GRB nature.

**Random probability of gamma-ray bursts coincidence with stars**

The number of coincident evens is $N = S_{coin} N_b N_{st} / \Omega = 0.18$, where $S_{coin} = \pi \sigma^2 = 0.04$ degree$^2$ – square of every event, $N_b = 391$, $N_{st} = 470$, $\Omega = 41253$ degree$^2$ – total sky surface. The Poisson probability is equal $W = e^{-N} N^4 / 4! = 4 \cdot 10^{-5}$.

This value is small enough what gives definite evidence for real stellar CGB identification.

Presented investigation gives evidence to gamma-ray bursts from nearby stars and also shows feather way for search of new stellar sources. It is no exclude that all not cosmological bursts (~ 60 %) are caused also by stars. The next task is to find method of discovery of active stars at greater distances (r > 20 ps). Certainly it is possible for detectors with higher angular resolution.

It is not worth to list the found stars due to its small number. This necessary study will be carried out when statistic significantly will grow and possibly give rise a new method to extract stars generating gamma-ray bursts. One has also to find differences between cosmological and stellar gamma-ray bursts.